\documentclass[sigconf,screen,nonacm]{acmart}

\usepackage{algorithmic}
\usepackage{graphicx}
\usepackage{subcaption}
\usepackage{xcolor} 
\usepackage{amsmath,amsfonts,amsthm}
\usepackage{braket}
\usepackage[ruled,vlined,linesnumbered]{algorithm2e}
\usepackage{graphicx}
\usepackage{textcomp}
\usepackage[normalem]{ulem}
\usepackage{xcolor}
\usepackage{xspace}
\usepackage{balance}
\usepackage{lineno} 
\usepackage{xurl}
\usepackage{tabularx}
\usepackage{multirow}
\usepackage{pifont}
\usepackage{enumerate}
\usepackage[shortlabels]{enumitem}
\usepackage{booktabs}
\usepackage{tcolorbox}
\usepackage{cleveref}
\usepackage{csquotes}

\newcommand{\tool}[1]{\textsc{#1}\xspace}

\newcommand{\mytool}{\tool{Codexity}}

\usepackage{multirow}

\usepackage{listings}
\makeatletter
\newcommand{\linebreakand}{%
  \end{@IEEEauthorhalign}
  \hfill\mbox{}\par
  \mbox{}\hfill\begin{@IEEEauthorhalign}
}
\makeatother

\begin{document}
\title{Codexity: Secure AI-assisted Code Generation}

\begin{abstract}
Despite the impressive performance of Large Language Models (LLMs) in software development activities, recent studies show the concern of introducing vulnerabilities into software codebase by AI programming assistants (e.g., Copilot, CodeWhisperer). In this work, we present \mytool, a security-focused code generation framework integrated with five LLMs. \mytool leverages the feedback of static analysis tools such as Infer and CppCheck to mitigate security vulnerabilities in LLM-generated programs. Our evaluation in a real-world benchmark with 751 automatically generated vulnerable subjects demonstrates \mytool can prevent 60\% of the vulnerabilities being exposed to the software developer.

\begin{itemize}[leftmargin=15pt]
\item Video: \url{https://youtu.be/om6CYGfbNnM}
\item Tool and Data:
\url{https://github.com/Codexity-APR/Codexity},
\url{https://doi.org/10.5281/zenodo.10572275}
\end{itemize}

\end{abstract}
\author{Sung Yong Kim}
\affiliation{%
  \institution{National University of Singapore}
  \country{Singapore}}
\email{sungyongkimmoi@gmail.com}

\author{Zhiyu Fan}
\affiliation{%
  \institution{National University of Singapore}
  \country{Singapore}}
\email{zhiyufan@comp.nus.edu.sg}

\author{Yannic Noller}
\affiliation{%
  \institution{Singapore University of Technology and Design}
  \country{Singapore}}
\email{yannic.noller@acm.org}

\author{Abhik Roychoudhury}
\affiliation{%
  \institution{National University of Singapore}
  \country{Singapore}}
\email{abhik@comp.nus.edu.sg}

\renewcommand{\shortauthors}{Kim et al.}

\maketitle

\section{Introduction}
\label{sec:intro}
The recent advance of fundamental Large Language Models (LLMs) for code has shown surprising results in many software engineering applications like code completion, code summarization, etc. Consequently, AI programming assistants like Copilot and CodeWhisperer have reshaped the way of software development. 
Despite their advantages and conveniences, researchers have expressed concerns that LLMs can introduce security vulnerabilities into the auto-generated code that developers could overlook ~\cite{asleep,fu2023security,khoury2023secure}. Without a secure integration mechanism for these LLM-generated codes, they could thus compromise the entire software system. Automated Program Repair (APR)~\cite{apr} is a technique that aims to fix bugs automatically, with many successful applications in fixing security vulnerabilities~\cite{zhang2022program,huang2019using,gao2021beyond,nguyen2013semfix,shariffdeen2021concolic}. However, existing APR approaches primarily focus on fixing human-written programs and typically assume timeouts of $\ge 1$ hour~\cite{trust_apr}. Such assumptions might fit in CI workflows or nightly build scenarios but do not satisfy the tight requirements for an \textit{in-time} vulnerability prevention needed for the automated code generation/completion inside the IDE.

To tackle these challenges, we present \mytool, the first security-focused code generation framework. \mytool integrates LLMs with static analyzers~\cite{CppCheck,infer2011} to establish security awareness and act as the first guard to prevent potential vulnerabilities introduced by AI programming assistants. In our experiments with 990 real-world code completion attempts, we demonstrate that, compared to ChatGPT, \mytool prevents the generation of 60\% of the vulnerabilities.
Our core contributions are as follows:
\begin{itemize}[leftmargin=15pt]

\item \mytool, a security-focused code generation framework providing timely vulnerability-free LLM-generated programs for developers. Additionally, \mytool employs a flexible architecture that allows user-customized LLM deployment.

\item Real-world benchmark for LLM-generated programs, consisting of 751 vulnerable subjects from 90 publicly available prompts.

\end{itemize}

\lstdefinestyle{customc}{
    belowcaptionskip=1\baselineskip,
    breaklines=true,
    frame=single,
    xleftmargin=\dimexpr\fboxsep-\fboxrule,
    xrightmargin=\dimexpr\fboxsep-\fboxrule,
    language=C,
    showstringspaces=false,
    basicstyle=\footnotesize\ttfamily,
    keywordstyle=\bfseries\color{black},
    commentstyle=\itshape\color{purple!40!black},
    identifierstyle=\color{black},
    stringstyle=\color{black},
    morekeywords={scanf},
    moredelim=[is][\color{red}]{|}{|}, 
    moredelim=[is][\color{green!40!black}]{*}{*}, 
}

\begin{figure*}[t]
\centering

\begin{minipage}[t]{0.45\textwidth}
\begin{lstlisting}[style=customc, caption={Vulnerable code generated by LLM}, label=lst:vul_code]
#include <stdio.h>
#include <errno.h>
#include <string.h>
int main(void){
  char reference[50], query[50]; 
  printf("\n Enter reference genome file name:  ");
  |scanf("%s", reference);|
  printf("\n Enter query genome file name:  ");
  |scanf("%s", query);| 
  return 0;
}
\end{lstlisting}
\end{minipage} \quad\quad\quad
\begin{minipage}[t]{0.45\textwidth}
\begin{lstlisting}[style=customc, caption={Safe code generated by LLM}, label=lst:secure_code]
#include <stdio.h>
#include <errno.h>
#include <string.h>
int main(void){
  char reference[50], query[50]; 
  printf("\n Enter reference genome file name:  ");
  *scanf("%49s", reference);* 
  printf("\n Enter query genome file name:  ");
  *scanf("%49s", query);* 
  return 0;
}
\end{lstlisting}
\end{minipage}
\end{figure*}

\section{Motivating Example}
\label{sec:motivating-example}

Listing~\ref{lst:vul_code} displays a program generated by an LLM. It is meant to answer a StackOverflow inquiry\footnote{\url{https://stackoverflow.com/questions/19063213/string-ask-and-store}} about storing user-input strings in variables. Our assembly of Static Application Security Testing (SAST) tools identified a potential vulnerability (highlighted in red in) that is related to CWE-119:

\begin{center}
"\textbf{CWE-119} identified at lines 7 and 9. The triggered concern:\\\textit{scanf() without field width limits can crash with huge input data}."
\end{center}
\begin{figure}
    \centering
    \includegraphics[width=0.9\linewidth]{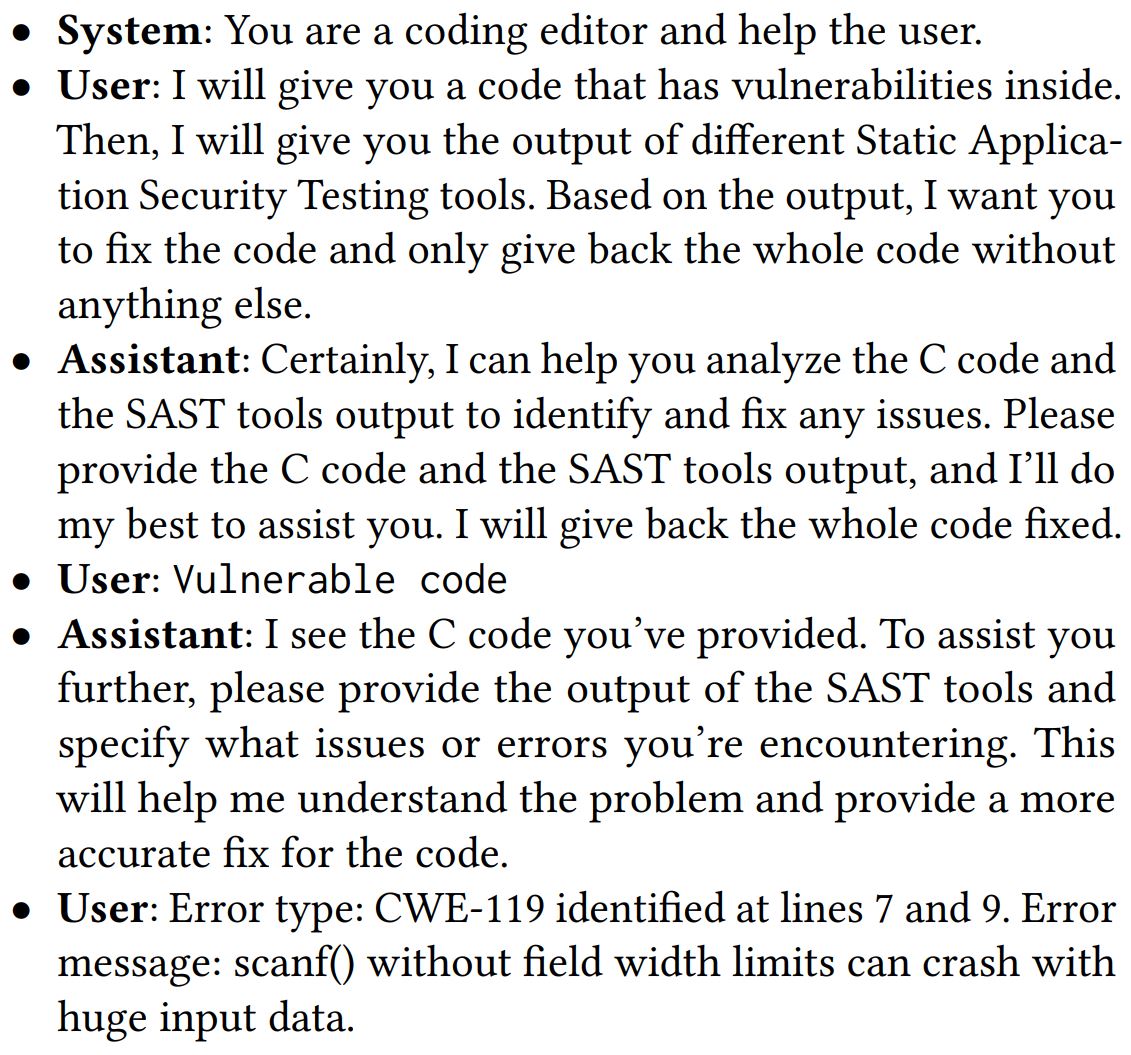}
    \caption{Conversational Prompt (\texttt{Vulnerable Code} is a placeholder for the code in Listing~\ref{lst:vul_code})}
    \label{fig:prompt}
\end{figure}
To prevent such vulnerabilities from being exposed to the software developer, we provide this additional feedback to the LLM in the form of a structured prompt, emulating a dialogue with the goal of crafting code devoid of these flaws (see Figure~\ref{fig:prompt}). Consequently, the LLM responded by producing the revised code, eliminating the buffer overflow issues, as evidenced in Listing~\ref{lst:secure_code}. As highlighted in green, the code is no longer vulnerable because the format specifier \%49s is used in both \texttt{scanf} calls, meaning that \texttt{scanf} will add at most 49 characters into the arrays. By specifying the maximum number of characters, the code prevents buffer overflow vulnerabilities if the input data exceeds the allocated buffer size.
\section{Design and Methodology}
\label{sec:methodology}


This section introduces the workflow of \mytool and two vulner\-ability-preventing repair strategies.

\subsection{\mytool Workflow}
In \mytool's workflow, the user first needs to select a repair strategy in the configuration setting to activate the system. \mytool currently offers the two strategies \textit{Iteration Repair} and \textit{Preshot Repair} catering for different requirements in the computation power. Then, the user can invoke \mytool to complete their code while programming. \mytool will take the existing code snippet to initiate a prompt and generate an initial completion with the selected LLM. The completed code will be routed to a vulnerability detection phase by a series of static analysis tools. \mytool is integrated with two state-of-the-art static analyzers CppCheck~\cite{CppCheck} and Infer~\cite{infer2011}. We selected these two tools to examine a wide range of vulnerabilities. Like Arusoaie et al.~\cite{8531281}, we also found CppCheck to be a generally good candidate. We added Infer because of its great ability to find memory-related bugs. If static analysis tools report any vulnerability, \mytool extracts the error/warning message and location information, along with the vulnerable program, to formulate a vulnerability-exposing prompt (see Figure~\ref{fig:prompt}). Finally, \mytool sends the vulnerability-exposing prompt to the LLM in the background and requests a vulnerable-free program. 





\begin{figure}[t]
    \centering
    \begin{subfigure}{0.4\linewidth}
        \includegraphics[width=\linewidth]{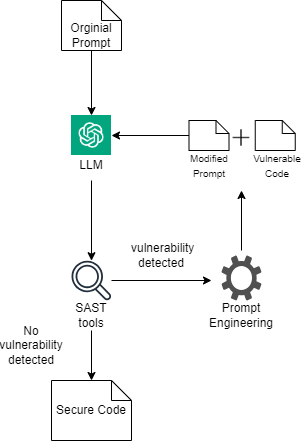}
        \caption{Iteration Repair}
        \label{fig:iteration}
    \end{subfigure}
    \hfill
    \begin{subfigure}{0.4\linewidth}
        \includegraphics[width=\linewidth]{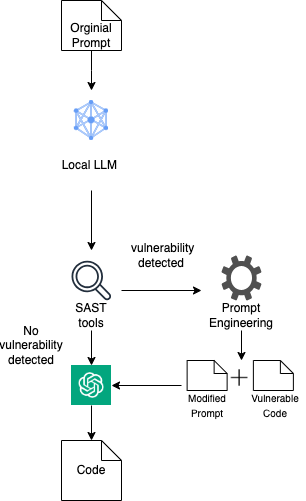}
        \caption{Preshot Repair}
        \label{fig:preshot}
    \end{subfigure}
    \caption{Codexity's Repair Strategies}
    \label{fig:strategies}
\end{figure}

\subsection{Strategy 1: Iteration Repair}

We now describe the default vulnerability-preventing strategy in \mytool. The \textit{Iteration Repair} strategy employs commercial LLMs to generate vulnerable-free programs and provides an interface that allows users to specify the exact background LLM engine. The goal is to leverage the most powerful LLM that the user can access to provide the most secure code generation service. Moreover, the API-accessible LLMs do not require users to be locally equipped with a high-end GPU for model inference. After generating a code, if a vulnerability is detected, the vulnerable code is sent back to the commercial LLM using its API. We repeat this until it generates one program for which the static analyzers report no vulnerability.








\subsection{Strategy 2: Preshot Repair}






Iteration repair aims to generate the best result by iteratively querying the model until a secure answer is found or the maximum allowed queries are reached. Although commercial LLMs provide better results, they might introduce high costs to the users. To balance the trade-off between cost and quality, we present an alternative vulnerability-preventing strategy called \textit{Preshot Repair} shown in Figure~\ref{fig:preshot}.
We hypothesize that a weaker local (and hence cheaper) LLM is more prone to generate vulnerabilities.
Such vulnerability information can be included in the query to the more costly and powerful LLM. 
Consequently, we aim to anticipate and prevent the flaws likely to surface in our target LLM's output, thus the term \textit{Preshot Repair}.
So, we first query a local LLM to generate an initial (vulnerable) code. We employ static analyzers to detect any vulnerability and, subsequently, use the analysis report in the prompt generation for the target LLM.

\section{Evaluation}
\label{sec:evaluation}

We show \mytool's effectiveness to securely handle developers' code completion requests based on two real-world datasets and compare \mytool's performance with FootPatch~\cite{footpatch_icse2018} and GitHub Copilot~\cite{copilot}. In particular, we explore three research questions:

\begin{description}[leftmargin=10pt]\itemsep0em 
    \item[RQ1:] How effective is \mytool in preventing the generation of vulnerable programs?
    \item[RQ2:] How does \mytool perform compared to FootPatch and GitHub Copilot?
    \item[RQ3:] What are the benefits/drawbacks of the proposed workflows? 
\end{description}

\subsection{Evaluation Setup}
\begin{figure}
    \centering
    \includegraphics[width=0.75\linewidth]{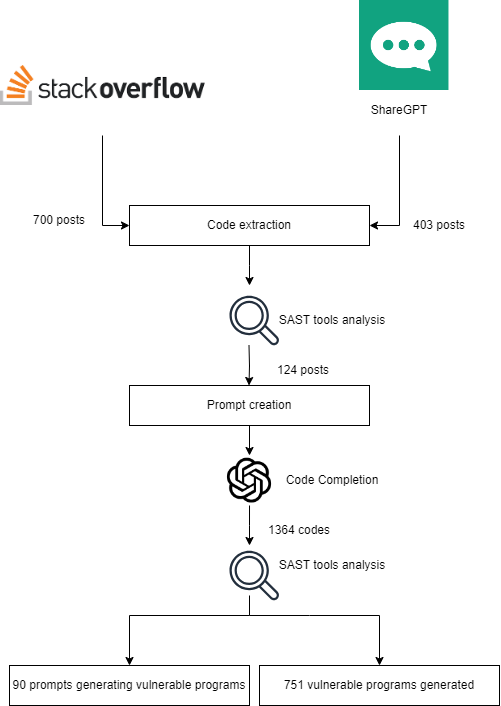}
    \caption{Dataset Construction}
    \label{fig:dataset}
\end{figure}
\subsubsection*{Benchmark Construction}

To simulate real-world scenarios of how AI-assisted programming may help software developers, we curated a benchmark of developer prompts that are prone to generate vulnerable programs from ShareGPT~\cite{ShareGPT} and StackOverflow~\cite{Stackoverflow}. ShareGPT is a platform where LLM users can post their experience in formulating prompts and interactions with LLMs, whereas StackOverflow is a popular programming-oriented Question-Answering platform. 
Figure~\ref{fig:dataset} shows the benchmark construction process.
We first collected 403 user posts relevant to C programming queries by filtering all the posts in the ShareGPT dataset and the first 700 posts in StackOverflow using \textit{`c'} and \textit{`int main'} as keywords. Each user post consists of a question and an LLM response. 
We then employ a two-round vulnerable prompts detection strategy. In the first round, we extracted the code snippet from the LLM response and ran the static analyzers in \mytool on the extracted code snippet of the 403 programming posts. As a result, we identified 124 posts containing vulnerabilities in their response. 

In the second round, we extracted the non-vulnerable part of the code snippet from the 124 posts and asked ChatGPT to complete the programming questions (using two temperature configurations) to confirm further whether they are prone to generate vulnerable programs. First, ChatGPT generates one completion of each prompt for the temperature at 0. Then, we executed ChatGPT again to generate ten completions of each prompt for temperature at 0.8. Finally, we reran the static analyzers on the 124$\times$11=1,364 LLM-generated code completions. We specified a prompt as \textit{vulnerable} if (1) ChatGPT produces a vulnerable program when the temperature is set to 0, or (2) ChatGPT produces at least one vulnerable program in one of the ten runs when the temperature is set to 0.8. Finally, we ended up with 90 vulnerable prompts and generated 751 vulnerable code completions with 1645 vulnerabilities.

Table~\ref{tab:dataset} shows the details of our benchmark, categorized by vulnerabilities identified by CppCheck and Infer. We employed CppCheck for common CWEs and Infer for specific types like Memory leaks. Note that subjects could have multiple vulnerabilities. We defaulted to CWE classification when vulnerabilities overlapped.

\begin{table}[!htb]
    \centering
    \caption{Vulnerabilities Categories}
    \begin{minipage}{.5\linewidth}
      \centering
        \begin{tabular}{|l | r|} 
            \hline
            \textbf{Vulnerability} & \textbf{Total} \\ [0.5ex] 
            \hline
            Null Dereference & 384\\
            Nullptr Dereference & 363\\
            Resource Leak & 196 \\
            Buffer Overrun L2 & 119\\
            Memory Leak & 109\\
            CWE-119 & 90\\
            CWE-457 & 81\\
            CWE-401 & 66\\
            Buffer Overrun L3 & 62\\
            CWE-775 & 32\\
            CWE-788 & 22 \\
            Buffer Overrun S2 & 15\\
            \hline
        \end{tabular}
    \end{minipage}%
    \hfill
    \begin{minipage}{.5\linewidth}
      \centering
        \begin{tabular}{|l | r|} 
            \hline
            \textbf{Vulnerability} & \textbf{Total} \\ [0.5ex] 
            \hline
            CWE-686 & 15\\
            Integer Overflow L2 & 13\\
            CWE-758 & 12\\
            CWE-197 & 12\\
            CWE-562 & 11\\
            Buffer Overrun L1 & 11\\
            CWE-467 & 10\\
            Use After Lifetime & 8\\
            CWE-685 & 7\\
            CWE-476 & 4\\
            Use After Free & 2\\
            Inferbo Alloc Is Zero & 1\\

            \hline
        \end{tabular}
    \end{minipage}
\label{tab:dataset} 
\end{table}


\subsubsection*{Baseline Tools}
We use FootPatch and GitHub Copilot as baseline tools in our comparison. We selected FootPatch because it also uses a SAST tool, Infer, to detect vulnerabilities. Also, we decided to compare with GitHub Copilot because it is one of the most famous code completion tools and runs with a similar GPT model.

\subsubsection*{Models}
Throughout the experiment, we used the latest ChatGPT (gpt-3.5-turbo) as a commercial LLM code completion and fixing engine for \mytool. For the preshot repair, we used two different specialized code completion models: StarCoder~\cite{li2023starcoder} in GPTQ~\cite{frantar2022gptq} format and SantaCoder~\cite{allal2023santacoder} in GGML~\cite{GGML} format. These formats allow quantization, suitable for CPUs or less powerful GPUs. The models were run with a maximum prediction token of 1024 to be able to complete the whole code and with a temperature of 0.2 (following the same evaluation setup as the authors of StarCoder). For the GPTQ model, we used the publicly available model at the Hugging Face repository\footnote{\url{https://huggingface.co/TheBloke/starcoder-GPTQ}}. Hugging Face is the equivalent of GitHub for sharing models. We obtained the GGML model by following the instructions in their GitHub repository.

\begin{table*}
    \centering
        \caption{Number of Vulnerable Codes generated}
    \begin{tabular}{|l|c|c|c|c|} \hline 
         & \textbf{\# of vulnerable codes} & \textbf{\% of vulnerable codes}& \textbf{Avg Time (s)} & \textbf{LoC} \\ \hline 
         \textbf{ChatGPT} & 751/990 & 75.9 & 39.2s& 55.2 \\ \hline 
         \textbf{IR-ChatGPT} & 157/990 & 15.9& 82.8s& 60.6 \\ \hline 
         \textbf{PR-StarCoder-15.5B} & 389/990 &  39.3& 45.4s& 54.9 \\ \hline 
         \textbf{PR-SantaCoder-1.1B}& 459/990 & 46.4& 72.4s& 44.4 \\ \hline
    \end{tabular}

    \label{tab:results}

    
\end{table*}

\subsubsection*{Technical Setup}
Our experiments were conducted on a MacBook Pro 2019 (Intel Core i9 with 16GB RAM) for iteration repair and preshot repair with the GGML model. For the GPTQ model in the preshot strategy, we used a Ubuntu 18.04 Server with RTX 4090.


\subsection{Experiment Result}

\subsubsection*{RQ1: Vulnerability Prevention}
We evaluated our tool on our 90 prompts that led to a vulnerable code generation and measured the number of vulnerable codes generated. We initially set a temperature of 0 for repairs in the iteration repair, as a high temperature can lead to a random output as mentioned in OpenAI documentation\footnote{\url{https://platform.openai.com/docs/api-reference/chat/create}}. A code is determined vulnerable when the SAST tools detect a vulnerability. We also set the maximum number of iterations to 3 for the iteration repair. Table~\ref{tab:results} shows the results of using \mytool in completing the \textit{vulnerable} prompts of our benchmark. As mentioned in Section~\ref{dataset_collection}, the baseline tool ChatGPT generated for the 990 code completion attempts in our dataset 751 vulnerable programs (751/990=75.9\%) with 1645 detected vulnerabilities. In the iteration repair setting, \mytool only generated 157 (157/990=15.9\%) vulnerable programs with 338 detected vulnerabilities. Therefore, it achieves a reduction of 60\% vulnerable programs.
%
%
The results in Table \ref{tab:results} also show that preshot repair decreases the production of vulnerable codes by 36.6\% when using StarCoder and 29.5\% for SantaCoder. In some cases (6.1\% for StarCoder and 8.5\% for SantaCoder), the tool did not output a code but a comment, e.g., asking a question about the prompt. In this case, because it is impossible to say if the code is secure, we considered the output vulnerable.

\subsubsection*{RQ2: Comparison with FootPatch and GitHub Copilot}
We compare \mytool with FootPatch on the 751 vulnerable subjects detected by Infer.
%
%
FootPatch's Infer found 20 vulnerabilities, including 19 null dereferences and one memory leak but patched none. The analysis of the log data revealed that FootPatch failed to address Null Dereferences when capturing the variable name:
\begin{lstlisting}[language=C]
Found error: NULL_DEREFERENCE
[+] Patchable error: [vulnerable_code.c]:[96] :[pointer newNode last assigned on line 95 could be null and is dereferenced at line 96, column 9]
[+] Patch generation routine started for bug "NULL_DEREF".
[+] Looking for pvar last in pname addSpecific
[=] I found these typs for pvar last
[-] No type for pvar found
\end{lstlisting}
Instead of extracting the pointer "newNode", FootPatch extracted "last", leading to an incorrect variable identification and a subsequent failure in the patch development. Similarly, Infer identified a vulnerability for the category Memory Leak but omitted the variable's name, complicating the patch process. This behavior of FootPatch is described in \cite{footpatch_icse2018} as where the tool failed to resolve the program variable.
We also compared \mytool with GitHub Copilot using its chat functionality. We asked Copilot to "write the next lines of the code" on our 90 prompts and took the first answer. Infer and CppCheck found 76 (84.44\%) programs to be vulnerable.


\subsubsection*{RQ3: Tradeoffs}
The iteration repair avoids generating vulnerabilities with a high success rate but requires multiple commercial LLM calls, increasing costs and generation time. On average, there are 1.8 repair iterations when a vulnerability is detected, and a generation takes 98.6 seconds (see Table \ref{tab:results}). This can be explained by the introduction of new vulnerabilities after the iteration by the LLM, leading to additional iterations. On the other hand, preshot repair utilizes a local LLM, minimizing API calls and time. For the GPTQ model, the repair adds about 6 seconds. For the GGML model, the generation time is slower because the model is run on a CPU. Moreover, the vulnerable code generation rate is higher than the iteration repair for all the models because we do not give a second chance to the preshot repair in case a vulnerability is detected after the commercial LLM generation.
The iterative repair strategy is optimal for users prioritizing secure code generation, albeit with a longer generation time. Conversely, the preshot repair strategy benefits those needing swift code generation but with a heightened level of security. For instance, preshot repair using ChatGPT coupled with StarCoder offers an average generation time similar to ChatGPT yet provides improved security.

\section{Related Work}
\label{sec:related}

He et al.~\cite{he2023large} introduced an innovative learning strategy for improved secure code generation. Other research includes works by Olausson et al.~\cite{olausson2023demystifying}, Pearce et al.~\cite{zero}, and Charalambous et al.~\cite{BMC}, each exploring LLMs efficacy in fixing code vulnerabilities.
However, these studies often adopt vulnerability detection techniques that are not suitable for IDE-integrated code completion tasks.
Pearce et al.~\cite{zero} used CodeQL, a vulnerability detection tool that scales to large codebases but is inefficient for analyzing short, single code snippets.
Charalambous et al.~\cite{BMC} implemented the Bounded Model Checking method, limiting the vulnerability spectrum.
Conversely, Olausson et al.~\cite{olausson2023demystifying} chose to engage the model for analyzing the error message; adding an extra call to a model can increase the required detection time.
Chen et al.~\cite{chen2023frugalgpt} discussed using multiple LLMs to boost result quality.
In contrast to these existing works, we integrate LLMs and SAST tools, specifically for tackling vulnerabilities, while considering the efficiency and practicality of the code completion tasks in the IDE.
\section{Conclusion}
\label{sec:Conclusion}

We introduced \mytool that addresses vulnerabilities in LLM-generated code. It deploys two distinct repair strategies: iteration repair and preshot repair. Our evaluation shows that they can reduce the generation of vulnerable code by 60\%. In future, it will be essential to improve the code generation efficiency, e.g., by combining \mytool and LLM fine-tuning to reduce the need for repair iterations. Furthermore, it will be interesting to explore \mytool for other popular programming languages, such as Java or Python.




\begin{acks}

This work was partially supported by a Singapore Ministry of Education (MoE) Tier 3 grant "Automated Program Repair", MOE-MOET32021-0001.

\end{acks}

\bibliographystyle{ACM-Reference-Format}
\bibliography{main}





\appendix
\newpage

\begin{figure}[t!]
\centering
\captionsetup[subfigure]{justification=centering}
\begin{subfigure}{.45\textwidth}
    \centering
    \includegraphics[width=1\linewidth]{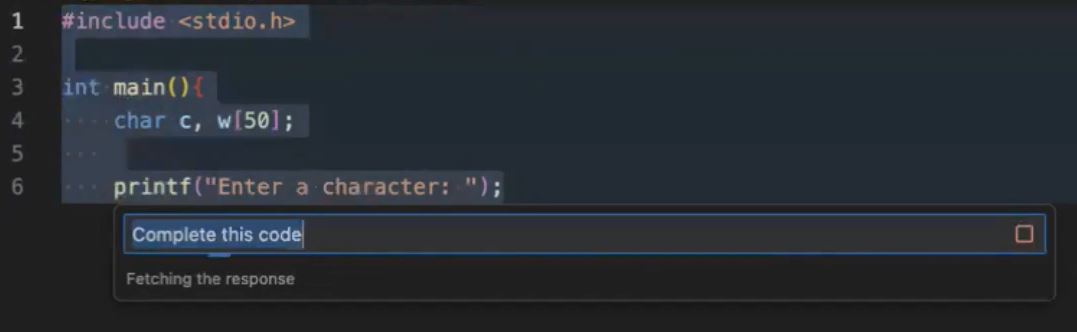}
    \caption{User Prompt}
    \label{fig:copilot_prompt}
\end{subfigure}
\begin{subfigure}{.45\textwidth}
    \centering
    \includegraphics[width=1\linewidth]{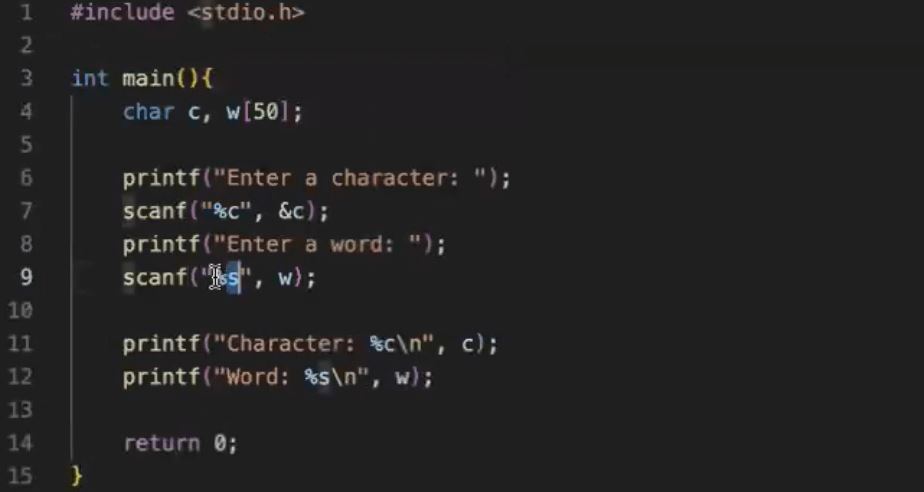}
    \caption{Code generated by GitHub Copilot}
    \label{fig:copilot_generated}
\end{subfigure}
\begin{subfigure}{.45\textwidth}
    \centering
    \includegraphics[width=1\linewidth]{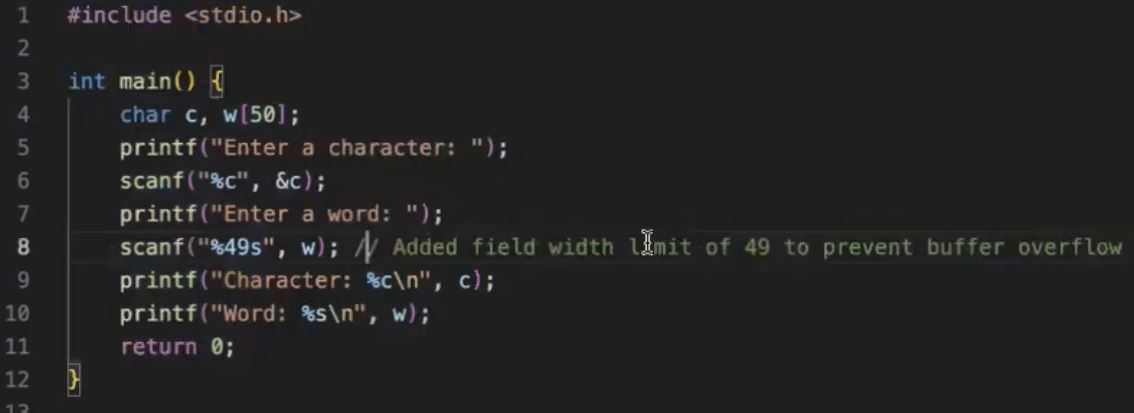}
    \caption{Code generated by \mytool with Iteration Repair}
    \label{fig:iteration_generated}
\end{subfigure}
\begin{subfigure}{.45\textwidth}
    \centering
    \includegraphics[width=1\linewidth]{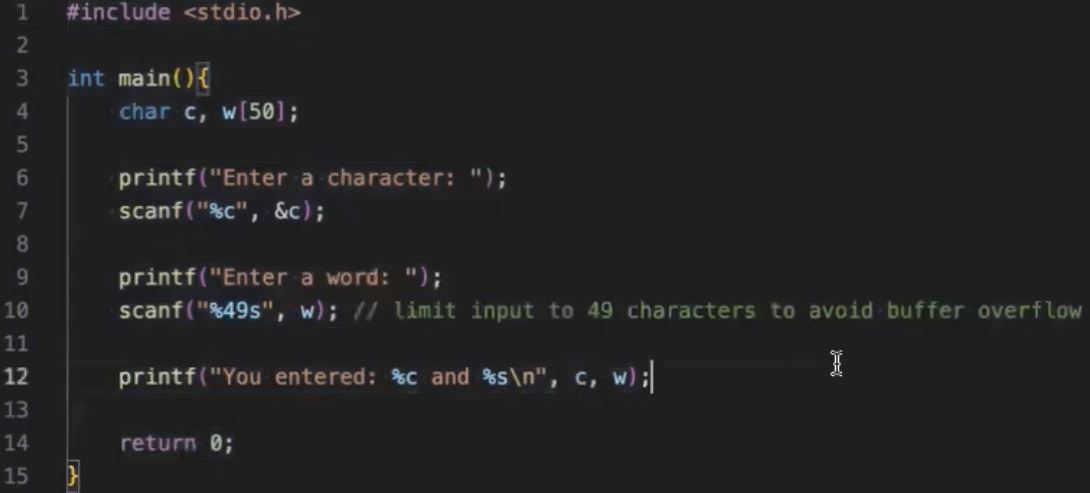}
    \caption{Code generated by \mytool with Preshot Repair}
    \label{fig:preshot_generated}
\end{subfigure}
\caption{Demo Screenshots}
\label{FIGURE LABEL}
\end{figure}

\section{Demo Walkthrough}
The demonstration of our tool is showcased in our video\footnote{\url{https://www.youtube.com/watch?v=om6CYGfbNnM}}, where we discuss the motivation, various strategies, and the practical use of \mytool. \mytool's demonstration involves the code completion for the snippet depicted in Figure \ref{fig:copilot_prompt}. We incorporated our tool into a Visual Studio Code extension for practical usage. In our demonstration, we showcase three ways of automated code completion in Visual Studio Code:

\begin{enumerate}
    \item utilizing GitHub Copilot,
    \item utilizing \mytool with the Iteration Repair strategy, and
    \item utilizing \mytool with the Preshot Repair strategy.
\end{enumerate}

Launching \mytool is similar to GitHub Copilot. The user needs to highlight the prompt that should be completed (e.g., a code comment or some partial code snippet) and then launch the extension, which will trigger our guided code completion using ChatGPT in the background.
We conclude the demonstration by comparing the outcomes of the three generated code variants. The code generated by GitHub Copilot exhibited a potential buffer overflow issue, as illustrated in Figure \ref{fig:copilot_generated} at line 9, due to the possibility of exceeding the maximum size of the variable 'w' based on the input. However, the iteration and preshot repair methods produce code that does not contain vulnerabilities (see Figures \ref{fig:iteration_generated} and \ref{fig:preshot_generated}).

\end{document}